%
\let\useblackboard=\iftrue
%
%
\newfam\black
\input harvmac.tex
\def\Title#1#2{\rightline{#1}
\ifx\answ\bigans\nopagenumbers\pageno0\vskip1in%
\baselineskip 15pt plus 1pt minus 1pt
\else
\def\listrefs{\footatend\vskip 1in\immediate\closeout\rfile\writestoppt
\baselineskip=14pt\centerline{{\bf References}}\bigskip{\frenchspacing%
\parindent=20pt\escapechar=` \input
refs.tmp\vfill\eject}\nonfrenchspacing}
\pageno1\vskip.8in\fi \centerline{\titlefont #2}\vskip .5in}

\ifx\answ\bigans\def\tcbreak#1{}\else\def\tcbreak#1{\cr&{#1}}\fi
\useblackboard
\message{If you do not have msbm (blackboard bold) fonts,}
\message{change the option at the top of the tex file.}
\font\blackboard=msbm10 scaled \magstep1
\font\blackboards=msbm7
\font\blackboardss=msbm5
\textfont\black=\blackboard
\scriptfont\black=\blackboards
\scriptscriptfont\black=\blackboardss

\else

\fi
%
\def\yboxit#1#2{\vbox{\hrule height #1 \hbox{\vrule width #1
\vbox{#2}\vrule width #1 }\hrule height #1 }}
\def\fillbox#1{\hbox to #1{\vbox to #1{\vfil}\hfil}}
\def\ybox{{\lower 1.3pt \yboxit{0.4pt}{\fillbox{8pt}}\hskip-0.2pt}}
\def\np#1#2#3{Nucl. Phys. {\bf B#1} (#2) #3}

\def\comments#1{}

\def\CA{{\cal A}}

\def\CF{{\cal F}}
\def\CT{{\cal T}}

\def\CN{{\cal N}}

\def\II{\relax{I\kern-.07em I}}

\def\IZ{\relax\ifmmode\mathchoice
{\hbox{\cmss Z\kern-.4em Z}}{\hbox{\cmss Z\kern-.4em Z}}
{\lower.9pt\hbox{\cmsss Z\kern-.4em Z}}
{\lower1.2pt\hbox{\cmsss Z\kern-.4em Z}}\else{\cmss Z\kern-.4em
Z}\fi}
\def\IB{\relax{\rm I\kern-.18em B}}
\def\IC{{\relax\hbox{$\inbar\kern-.3em{\rm C}$}}}
\def\ID{\relax{\rm I\kern-.18em D}}
\def\IE{\relax{\rm I\kern-.18em E}}
\def\IF{\relax{\rm I\kern-.18em F}}
\def\IG{\relax\hbox{$\inbar\kern-.3em{\rm G}$}}
\def\IGa{\relax\hbox{${\rm I}\kern-.18em\Gamma$}}
\def\IH{\relax{\rm I\kern-.18em H}}
\def\II{\relax{\rm I\kern-.18em I}}
\def\IK{\relax{\rm I\kern-.18em K}}
\def\IP{\relax{\rm I\kern-.18em P}}

\font\cmss=cmss10 \font\cmsss=cmss10 at 7pt
\def\IR{\relax{\rm I\kern-.18em R}}

\Title{ \vbox{\baselineskip12pt\hbox{hep-th/9608111}
\hbox{RU-96-69}}}
{\vbox{
\centerline{Five Dimensional SUSY Field Theories,}
\centerline{Non-trivial Fixed Points}
\centerline{and String Dynamics}}} 
\centerline{Nathan Seiberg}
\smallskip
\smallskip
\centerline{Department of Physics and Astronomy}
\centerline{Rutgers University }
\centerline{Piscataway, NJ 08855-0849}
\centerline{\tt seiberg@physics.rutgers.edu}
\bigskip
\bigskip
\noindent
We study (non-renormalizable) five dimensional supersymmetric field
theories.  The theories are parametrized by quark masses and a gauge
coupling.  We derive the metric on the Coulomb branch exactly.  We
use stringy considerations to learn about new non-trivial interacting
field theories with exceptional global symmetry $E_n$ ($E_8$, $E_7$,
$E_6$, $E_5=Spin(10)$, $E_4=SU(5)$, $E_3=SU(3)\times SU(2)$,
$E_2=SU(2)\times U(1)$ and $E_1=SU(2)$).  Their Coulomb branch is ${\bf
R}^+$ and their Higgs branch is isomorphic to the moduli space of $E_n$
instantons.  One of the relevant operators of these theories leads to a
flow to $SU(2)$ gauge theories with $N_f=n-1$ flavors.  In terms of
these $SU(2)$ IR theories this relevant parameter is the inverse gauge
coupling constant.  Other relevant operators (which become quark masses
after flowing to the $SU(2)$ theories) lead to flows between them.  Upon
further compactifications to four and three dimensions we find new fixed
points with exceptional symmetries.

\Date{August 1996}

\newsec{Introduction}

\nref\nclp{For a nice review see, S. Chaudhuri, C. Johnson, and J.
Polchinski, ``Notes on D-Branes,'' hep-th/9602052.}%
\nref\dgdb{M.~R.~Douglas, ``Gauge Fields and D-branes,'' hep-th/9604198.}%
\nref\bds{T. Banks, M.R. Douglas and N. Seiberg, ``Probing $F$-theory
With Branes,'' hep-th/9605199.}%
\nref\threedone{N. Seiberg, ``IR Dynamics on Branes and Space-Time
Geometry,'' hep-th/9606017.}%
\nref\threedtwo{N. Seiberg and E. Witten, ``Gauge Dynamics and
Compactification to Three Dimensions,'' hep-th/9607163.}%
\nref\dkps{M.~R.~Douglas, D.~Kabat, P.~Pouliot and S.~H.~Shenker,
``D-branes and Short Distances in String Theory,'' hep-th/9608024.}%

It is becoming clear that there is an interesting relation between
quantum field theory on branes \nclp\ and space-time dynamics in string
theory \refs{\nclp - \dkps}.  This relation generalizes the more
standard world-sheet/space-time relation.  It is particularly
interesting when the theory on the brane is a non-trivial quantum
theory.  In this context the work of \bds\ used non-trivial dynamics on
3-branes to explain the space-time results of
\ref\sen{A.~Sen, ``F-theory and Orientifolds,'' hep-th/9605150.}.
\nref\intse{K. Intriligator and N. Seiberg, ``Mirror Symmetry in Three
Dimensional Gauge Theories,'' hep-th/9607207.}%
The extension of these ideas to 2-branes has led to new results in
three dimensional quantum field theory \refs{\threedone, \threedtwo,
\intse}.  Here we continue this line of investigation by considering
4-branes.

Superficially, the dynamics on a 4-brane cannot be interesting.  The
relevant field theory is five dimensional and is believed to be
``trivial.''  However, we will show that already at one loop order some
interesting effects take place.  Furthermore, we will argue that certain
strongly coupled non-trivial fixed points exist.  They exhibit
exceptional global symmetries: $E_8$, $E_7$, $E_6$, $E_5=Spin(10)$,
$E_4=SU(5)$, $E_3=SU(3)\times SU(2)$, $E_2=SU(2)\times U(1)$ and
$E_1=SU(2)$.

In section 2 we review some basic facts about supersymmetric field
theories in five dimensions.  In section 3 we specialize to $U(1)$ and
$SU(2)$ gauge theories with various numbers of flavors and study them in
perturbation theory. In section 4 we use these five dimensional field
theories in string theory.  Our considerations lead us to the new fixed
points with $E_n$ symmetry.  They are further studied in section 5 where
we relate them to similar strongly coupled fixed points in fewer
dimensions.

\newsec{Review of Five dimensional SUSY}

The spinor representation of $SO(4,1)$ is four dimensional and is
pseudoreal.  Since the vector of $SO(4,1)$ is in the antisymmetric
product of two spinors, the minimal SUSY algebra is generated by two
charges.  (Extended supersymmetry algebras also exist but will not be
discussed here.)  It is related by dimensional reduction to the minimal
SUSY algebra in six dimensions, to the more familiar $N=2$ algebra in
four dimensions and to $N=4$ SUSY in three dimensions.  As in all these
theories, an important role is played by the $SU(2)_R$ automorphism of
this algebra under which the two supercharges are a doublet.

The massless representations of this algebra are the hypermultiplet
(four real scalars and a spinor) and a vector (a vector, a real scalar
and a spinor).  The vector representation is equivalent under duality to
a tensor representation (a two form, a real scalar and a spinor).
Therefore, if we start {}from six dimensions the vector and the tensor
become isomorphic representations in five dimensions.

All these theories have Coulomb branches which are parametrized by the
expectation values of the scalars $\phi^i$ in vector multiplets. There
can also be Higgs branches where the hypermultiplets vary.  They are
hyper-Kahler manifolds.  The most general Lagrangian (with up to two
derivatives) on the Coulomb branch is easily determined.  If we reduce
to four dimensions, it should satisfy special geometry and hence it is
derived {}from a prepotential\foot{Since the $\CA^i$ are real in five
dimensions, we find it convenient to redefine the prepotential by a
factor of $i$ compared with the standard four dimensional definition.}
$\CF(\CA^i)$ which is locally a function of the vector superfields
$\CA^i$.  The requirements of five dimensional SUSY can be implemented
as follows.  In the reduction to four dimensions the fifth components of
the vectors $A_5^i$ become scalars thus making $\phi^i$ complex.
Invariance under $A_5^i \rightarrow A_5^i + a^i$ with $a^i$ arbitrary
real constants translates to invariance under $\CA^i
\rightarrow \CA^i + i a^i$.  This fixes $\CF$ to be at most cubic
\eqn\cfinf{\CF= c_0 + c_i \CA^i + c_{ij} \CA^i \CA^j + c_{ijk} \CA^i
\CA^j \CA^k.}
The constants $c_0$ and $c_i$ do not affect the Lagrangian and can be
set to zero.  The reality properties of the Lagrangian and the
invariance mentioned above restrict the constants $c_{ij}$ and $
c_{ijk}$ to be real.

In six dimensions there are both tensor and vector multiplets.  It is
not known how to write Lagrangians with an arbitrary number of tensor
multiplets.  If there is only one tensor multiplet $\CT$ and we add a
gravity multiplet, a Lagrangian can be written.  It is derived {}from a
prepotential $\CF= c_{ij} \CA^i \CA^j + \CT^2 + \tilde c_{ij} \CT \CA^i
\CA^j$ which becomes a special form of \cfinf\ upon reduction to five
dimensions.

Consider for simplicity the case of only one vector multiplet.  Then
\cfinf\ takes the form
\eqn\cfinfo{\CF= {1 \over 2 g^2} \CA^2 +  {c \over 6} \CA^3}
with the constants $g$ and $c$ being real.  The expression \cfinfo\ is
valid only locally (we will see this more explicitly below).  Therefore,
we will not use the freedom to shift $\CA$ by a constant to set the
quadratic term to zero.

In components the first term in \cfinfo\ yields the kinetic terms of the
fields in the multiplet whose coefficient $1 \over g^2$ depends on the
gauge coupling $g$.   The cubic term leads to terms proportional to
\eqn\intterm{c \left ( \phi (\partial \phi)^2 +  \phi F_{\mu\nu}^2 +
A \wedge F \wedge F + \cdots \right).}
It is amusing to note that the $A \wedge F \wedge F$ term is
reminiscent of a similar term in the eleven dimensional supergravity
Lagrangian.  If we ignore this term we can perform a duality
transformation similar to the one in four dimensions.  The vector
multiplet becomes a tensor multiplet which includes a two form
$B_{\mu\nu}$ gauge field and a scalar
\eqn\phidd{\phi_D={\partial \CF \over \partial \CA }(\phi)= {1\over
g^2} \phi + {c \over 2} \phi^2.}

The massive spectrum can include BPS saturated states.  Their masses are
given by the expectation values of $\phi$ and $\phi_D$.  Particle are
electrically charged and their masses are given by
\eqn\bpse{m/\sqrt 2 = Z_e = n_e (\phi + c_e)}
and strings are magnetically charged and their tensions are given by
\eqn\bpsm{T /\sqrt 2 = Z_m = n_m (\phi_D  + c_m).}
Note the freedom in shifting $\phi$ and $\phi_D$ by constants.
If there are also global Abelian symmetries, \bpse\ can receive other
contributions associated with the charges of these symmetries.

\newsec{Perturbative Dynamics}

In this section we study $U(1)$ gauge theories with $N_f$ ``electron''
hypermultiplets of charge one and $SU(2)$ gauge theories with $N_f$
``quark'' hypermultiplets ($2N_f$ half-hypermultiplets) in the two
dimensional representation.  The Coulomb branch of the moduli space of
the $U(1)$ theory is $\bf R$ while for the $SU(2)$ theory it is ${\bf
R}/{\bf Z}_2= {\bf R}^+$.  These field theories are not renormalizable.
Therefore, they should be viewed as field theories with a cutoff.  Even
if in the classical theory the cubic term in the prepotential vanishes,
it can be generated in the quantum theory
\ref\mandf{E. Witten, ``Phase Transitions In M-Theory And F-Theory,''
hep-th/9603150.}.
We will refer to this phenomenon as an anomaly.
Clearly, $c$ can be generated only at one loop (it is independent of
$g$).  As a finite quantity it is independent of the cutoff.  It is easy
to see that only chiral objects contribute to $c$.  Therefore, only
states in small representations can affect it and a contribution of a
hypermultiplet has the same absolute value but the opposite sign to that
of a vector multiplet.  Also, it is clear that the contribution of any
multiplet is proportional to the cube of its charge (the $CP$ conjugate
representation has the opposite charge but also the opposite chirality
and therefore contributes the same).  Therefore, for the $U(1)$ theory
$c= -a N_f $ while for the $SU(2)$ theory $c= a(8-N_f)$ for some
constant $a$ which can be determined by an explicit one loop
computation.  The sign of $a$ is important and turns out to be positive.
The relevant one loop computation was performed in \mandf.  We will
absorb $a$ in 
the normalization of the action and the gauge coupling $g$ and set
\eqn\cone{c= \cases{- N_f & for ~ $U(1)$ \cr
2(8-N_f) & for ~ $SU(2)$.\cr}}

This anomaly term is similar to the standard anomaly in four dimensions
and the anomaly discussed in \refs{\threedone,\threedtwo} in three
dimensions.  All of these receive contributions only {}from massive BPS
states with hypermultiplets and vector multiplets contribute with
opposite signs.  In the $U(1)$ theories all of these anomalies are
proportional to $N_f$.  Since in four dimensions the contribution of
each multiplet is proportional to the square of the charge and in three
dimensions it is proportional to the charge, in the $SU(2)$ theory the
anomaly is proportional to $2(4-N_f)$ in four dimensions and to
$2(2-N_f)$ in three dimensions \threedtwo.

The expression \cfinfo\ is valid only locally.  At various points in the
moduli space there can be singularities.  For example, in the $U(1)$
theory there is a singularity at $\phi=0$ where the electrons become
massless.  In order to extend the Lagrangian beyond that point we use
the global symmetry which acts on the scalar as $\phi \rightarrow -\phi$
combined with parity.  This is a symmetry of the underlying Lagrangian
(in six dimensions it is part of the Lorentz group).  The bosonic terms
\intterm\ are invariant only if they are
\eqn\intterme{c \left ( |\phi| (\partial \phi)^2 +  |\phi| F_{\mu\nu}^2
+ \epsilon(\phi) A \wedge F \wedge F + \cdots \right).}

We see that the effective gauge coupling in the $U(1)$ theory is 
\eqn\effg{{1\over g_{eff}^2} = {1\over g^2} + c |\phi| .}
It is continuous but not smooth at $\phi=0$.  The discontinuity in its
derivative is proportional to $c$.  Since $c$ is negative, for every
bare coupling $g$, the effective coupling $g_{eff}$ diverges at finite
points in the moduli space $\phi_s = \pm {1 \over c g^2}$.  This
divergence reflects the fact that such a quantum field theory is not
renormalizable and more data is needed at high energy, of order $1 \over
g^2$, to define it.

It is straightforward to generalize this result to a $U(1)$ gauge
theory with several massive electrons with masses $m_i$.  (Clearly, we
have the freedom to redefine the origin of $\phi$ and thus set one of
the masses to zero.)  The effective gauge coupling is
\eqn\effgs{{1\over g_{eff}^2} = {1\over g^2} - \sum_i |\phi-m_i|.}

In the $SU(2)$ theory, where the moduli space is modded out by $\phi
\rightarrow -\phi$, we take $\phi \ge 0$ and there is a singularity at
the end of the moduli space at $\phi=0$.  With several quarks with
masses $m_i$ the effective gauge coupling is
\eqn\effgs{{1\over g_{eff}^2} = {1\over g^2} + 16 \phi -\sum_i
|\phi-m_i| -\sum_i |\phi + m_i| ~.} 
Note that the singularity at $\phi=m_i >0 $ is the same as in the $U(1)$
theory with one electron.  This is consistent with the low energy theory
at that point being $U(1)$ with one electron.

As in the $U(1)$ theories, the $SU(2)$ theories with $N_f>8$ have
singularities in the moduli space reflecting the lack of
renormalizability of the field theory.  On the other hand, for $N_f \le
8$ there is no singularity.  This suggests that for $N_f<8$ we can
consider the strong coupling limit $g=\infty$ and find a sensible
theory.  Indeed, below we will describe this theory and will show that
upon a small perturbation by $1 \over g^2$ it flows to the $SU(2)$
theory.

\bigskip
\centerline{\it A Peculiar $U(1)$ Symmetry}

In five dimensions the current 
\eqn\newcurrent{j={}^*(F\wedge F)}
is always conserved and the theory has a global $U(1)_I$ symmetry.  Its
charge is the instanton number, $I$.  As in
\ref\nonren{N. Seiberg, ``Naturalness Versus Supersymmetric
Non-renormalization Theorems,'' Phys.Lett. {\bf B318} (1993) 469,
hep-ph/9309335.},
we introduce parameters as background fields.  In this case these are
gauge superfields
\ref\aps{P.C. Argyres, M.R. Plesser and N. Seiberg,
``The Moduli Space of Vacua of $N=2$ SUSY QCD and Duality in $N=1$ SUSY
QCD,'' hep-th/9603042.}
associated with gauging global symmetries.  Coupling the conserved current
\newcurrent\ to vector superfields we identify the scalar component of
this vector superfield, $m_0$, as the gauge coupling $m_0 \sim {1\over
g^2}$.  In six dimensions $1\over g^2$ is in a tensor multiplet and the
charged objects are strings.  In five dimensions it is in a vector
multiplet and the charged objects are particles whose mass is determined
by a BPS formula to be related to the expectation value of the
background scalar $m_0$.  In the vacuum of the $SU(2)$ theory with
$\phi=0$ these are the four dimensional instantons which appear as
particles (zero branes) in five dimensions\foot{These instantons have a
non-compact dilation zero mode.  Its quantization leads to a continuous
spectrum of particles.  The quantization of the fermion zero modes makes
them spinors of $Spin (2N_f)$.  The presence of massless charged fields
makes the interpretation of this spectrum confusing.}.  In the vacua
with $\phi \not=0$ the instantons tend to shrink and their detailed
properties depend on the short distance physics, which depends on the
way the theory is regularized.  In particular, there could be various
BPS bound states.  This dependence on unknown short distance physics is
another manifestation of the lack of renormalizability of the theory.

Related to the fact that the global $U(1)_I$ symmetry with charged
particles exists in five dimensions is the fact that only in $d=5$ is
the dimension of $m_0 \sim {1\over g^2}$ exactly one.  $m_0$ behaves
like the quarks mass terms which can also be thought of as background
vector superfields.  The induced cubic term $\CA^3$ in \cfinfo\ leads to
interesting consequences for this symmetry.  It leads to a coupling of
the corresponding current \newcurrent\ to the dynamical gauge field
$\CA$.  In other words, the instantons carry gauge charge!  One way to
see that is to note that locally we can change $m_0$ by shifting $\CA$.

The central charge in the supersymmetry algebra which determines the
masses of BPS states is a linear combination of all $U(1)$ symmetries
whether gauged or not.  Therefore, superficially
\eqn\centraln{Z^{(0)}=I_3\phi + I m_0}
where $I_3$ is the electric charge (the value in the Cartan subalgebra
in the $SU(2)$ theory) and $I$ is the instanton charge.  The mixing of
the two symmetries converts $m_0 \sim {1 \over g^2} \rightarrow {1 \over
g_{eff}^2}$ and therefore
\eqn\central{Z=(I_3 + c I)\phi + I  m_0. } 

We would also like to point out that in the $SU(2)$ theory there are
strings.  In four dimensions these theories have magnetic monopoles.
These classical configurations become strings in five dimensions.  Their
tension is controlled by the BPS formula \bpsm.

\newsec{String considerations}

One context in string theory where these theories are important is in
compactifications of the type I theory on $S^1$.  We use as a probe the
ten dimensional 5-brane and wrap it around the compact circle to produce
a 4-brane in nine dimensions.  After dualizing the circle, the theory
is the type I' compactified on $S^1/{\bf Z}_2$ with two orientifolds at
the ends \nclp.  Our 4-brane probe can be understood then as the
4-brane of the IIA theory localized at a point in $S^1/{\bf Z}_2$.
The space-time moduli in the type I language are the size of the
original $S^1$ and Wilson lines.  The latter are 16 phases.  After the
duality, the moduli are the the size of $S^1/{\bf Z}_2$ and the
locations of 16 background D8-branes \nclp.  The theory on the 4-brane
probe is an $SU(2)$ gauge theory in five dimensions (and a few 
free fields).  The modulus of this theory is the location in $S^1/{\bf
Z}_2$.  At the two endpoints the $SU(2)$ gauge symmetry is restored.  It
is broken to $U(1)$ in between the two points.  The $SU(2)$ gauge
theory is coupled to $N_f=16$ hypermultiplets.  Their masses, which are
parameters of the theory on the brane, are the locations of the sixteen
background D8-branes.

For a generic radius, special points in the space-time moduli space
occur when $N_f$ D8-branes are at the same point on $S^1/{\bf Z}_2$.
Then, on the 4-brane probe we get $U(1)$ with $N_f$ electrons.  Another
special point is when $N_f$ D8-branes coincide with an orientifold.
Then the theory on the 4-brane probe is $SU(2)$ with $N_f$ quarks.  The
global enhanced symmetry in these cases, $SU(N_f)$ and $SO(2N_f)$
respectively, are enhanced gauge symmetries in space-time.  The Higgs
branches which emanate {}from the singularities are the moduli spaces of
space-time instantons in these enhanced gauge symmetries.

\nref\dl{M. Douglas and M. Li, ``D-Brane Realization of $\CN=2$ Super
Yang-Mills Theory in Four Dimensions,'' hep-th/9604041.}

The preceding two paragraphs are completely analogous to the discussion
in \bds\ for 3-brane probes and in \threedone\ for 2-brane probes
which arise in similar fashions.  One aspect, which was not
stressed in \refs{\bds,\threedone}, is the space-time meaning of the
anomaly -- the constant $c$.  As we said above, this anomaly is similar
to the ordinary anomaly in four dimensions and the anomaly discussed in
\refs{\threedone,\threedtwo} in three dimensions.  All of them arise at
one loop.  The BPS particles which run in the loop originate in string
theory {}from strings stretched between the probe and the orientifolds
(vector multiplets) and D-branes (hypermultiplets).  In string theory
this diagram is an annulus diagram.  As explained in \refs{\nclp,\dl},
this diagram can be viewed in the cross channel as a tree level exchange
between the probe and the orientifold or the background D-branes.  In
space-time, this tree level effect leads to a velocity dependent force
between our probe and the background orientifolds and D-branes.  These
background branes are charged under some space-time gauge symmetries.
In units where the charge of every background D-brane is -1, the charge
of the orientifold is 8 in nine dimensions, 4 in eight dimensions and 2
in seven dimensions.  Therefore, the charge of $N_f$ background D-branes
is $-N_f$ and the charge of $N_f$ background D-branes coinciding with an
orientifold is $8-N_f$ in nine dimensions, $4-N_f$ in eight dimensions
and $2-N_f $ in seven dimensions.  This is precisely the value of the
anomaly in the theory on the various probes we used -- 4-brane in nine
dimensions, 3-brane in eight dimension \bds\ and 2-brane in seven
dimensions \threedone\ (up to a trivial factor of two at the
orientifolds).

The global $U(1)_I$ symmetry associated with instanton number on the brane
also has an obvious interpretation in this space-time description.  Like
every global symmetry on the brane, this is a gauge symmetry in
space-time.  In order to identify this symmetry, recall that in the type
I theory the gauge coupling in the theory on the probe satisfies 
\eqn\typeIcou{{1\over g^2} \sim {R M_s^2\over \lambda_I} \sim {M_s \over
\lambda_{I'}}}
where $\lambda_I $ ($\lambda_{I'}$) is the dimensionless type I (type
I') coupling constant, $R$ is the radius of $S^1$ and $M_s$ is the
string scale.  Therefore, changing $R$ for fixed $\lambda_I$ amounts to
changing the gauge coupling.  The corresponding gauge field in
space-time is the $U(1)$ superpartner of the space-time modulus $R$.  A
simple check of this identification is the following.  An instanton on
our 4-brane is a particle.  In the underlying type I theory, this is
an instanton on the 5-brane and therefore it is a string.  This
string was identified in
\ref\dbwb{M.~R.~Douglas, ``Branes within Branes,'' hep-th/9512077.}
as the heterotic string.  In our case it wraps the compact $S^1$.
Therefore, it is charged under the $U(1)$ of winding number of the dual
heterotic string.

Consider now a configuration of background D8-branes such that $n_L$ of
them are at one orientifold, at $\phi=0$, $p$ of them are between the two
orientifolds at $\phi_i$ ($i=1,...,p$) and $n_R=16-p-n_L$ coincide
with the other orientifold at $\phi={1 \over R} $.  Let us probe this
system with our D4-brane probe.  Near $\phi=0$ the theory on the probe
is $SU(2)$ with $n_L$ flavors.  Near $\phi_i$ it is $U(1)$ with a
massless electron.  We could also think about it as $SU(2)$ with $n_L+p$
flavors, $p$ of them have masses $\phi_i$.  The effective gauge
coupling is 
\eqn\effcou{{1\over g^2_{eff}(\phi)} = \cases{ 
{1\over g^2(0)} + 2(8-n_L)\phi & for~ $0<\phi<\phi_1$ \cr
{1\over g^2_{eff}(\phi_i)} + 2(8-n_L-i)(\phi-\phi_i) & for~
$\phi_i<\phi<\phi_{i+1} $\cr
{1\over g^2_{eff}(\phi_p)} + 2(8-n_L-p)(\phi-\phi_p) & for~
$\phi_p<\phi<{1 \over R}$ ~. \cr}}
Using \typeIcou\ this change in the gauge coupling on the 4-brane probe
as a function of the moduli translates to a space dependent dilaton
field $\lambda_{I'}(\phi)$ in space-time.  This variation has already been
observed in
\ref\polwit{J. Polchinski and E. Witten, `` Evidence for Heterotic --
Type I Duality,'' \np{460}{1996}{525}, hep-th/9510169.}.
For sufficiently large $R$ the effective gauge coupling
$g^2_{eff}(\phi)$ and therefore also $\lambda_{I'}(\phi)$ remain finite.
However, at some $R_0$ which depends on $g(0)$, $n_L$ and $\phi_i$ they
diverge.  For example, for $p=0$ and $n_L>8$ they diverge when 
\eqn\conddi{{1\over g^2_{eff}(\phi={1 \over R_0})}= {1\over
g^2(0)} + 2(8-n_L){ 1 \over R_0} =0}
or when
\eqn\conddis{  {1 \over \lambda_{I'}(0)} \sim n_L- 8.}
We express it in terms of the dual heterotic string variables by using
$R_0^2=\lambda_IR_{0h}^2$, ($R_{0h}$ is the corresponding
heterotic radius) to find the condition ${R_{0h}^2 M_s^2}\sim n_L- 8 $ 
for enhancement of the gauge symmetry in the heterotic theory.  This
derivation parallels the space-time analysis of \refs{\polwit,\nclp}.
It extends it by showing that the result is exact; i.e.\ there are no
instanton corrections similar to those in \sen.

This computation for $p=0$ and $n_L>8$ is easily generalized to $p$
branes at $\phi_i$.  As we make $R$ smaller, the divergence first occurs
at an orientifold (which we will take to be the right one) with $n_R<8
$.  Let us consider the theory on the 4-brane probe for finite $g_0=
g_{eff}({1\over R})$ near the orientifold at $\phi= {1 \over R}$.  It is
an $SU(2)$ theory with $N_f=n_R<8$ flavors.  Its modulus is $\phi_R= {1
\over R}-\phi \ge 0 $ and its effective coupling is
\eqn\conddir{{1\over g^2_{eff}(\phi_R)}= {1\over g_0^2} +
2(8-N_f)\phi_R. } 

The long distance dynamics of the theory on the brane is always a local
quantum field theory.  Therefore, even as $g_0$ diverges, the theory on
the brane must make sense as a quantum field theory.  It must be at a
fixed point of the renormalization group.  As $g_0$ diverges, the
string theory acquires an enhanced gauge symmetry.  Correspondingly, the
theory on the brane should acquire enhanced global symmetry.  Therefore,
{\it for $g_0=\infty$ the theory on the brane is at a non-trivial fixed
point with enhanced global symmetry}.

For finite $g_0$ the symmetry of the five dimensional theory is
$SO(2N_f) \times U(1)$.  We will argue that the fixed point
corresponding to $g_0=\infty$ has an enhanced $E_{N_f+1}$ symmetry
($E_8$, $E_7$, $E_6$, $E_5=Spin(10)$, $E_4=SU(5)$, $E_3=SU(3)\times
SU(2)$, $E_2=SU(2)\times U(1)$ and $E_1=SU(2)$).  We will do that by
examining the symmetries of the underlying string theory.

When the $SO(32)$ heterotic string is compactified on $S^1$ an $E_8$
symmetry is obtained as follows.  An $SO(14)\times U(1)$ subgroup of
$SO(32)$ is combined with the left moving $U(1)$ associated with the
$S^1$.  For special values of the Wilson lines and the radius, this
$SO(14)\times U(1)\times U(1)$ is enhanced to $E_8\times U(1)$.  The new
gauge bosons which become massless at that point are winding modes of
the heterotic string.  Upon heterotic/type I/type I' duality a similar
enhancement is expected in the type I' theory \nclp.  It occurs at a
point where the type I' coupling $\lambda_{I'}$ diverges at one of the
orientifolds (if it diverges at both orientifolds the symmetry is
enhanced to $E_8\times E_8$).  This is exactly the phenomenon we
discussed above.  Therefore, for $N_f=7$ the strongly coupled theory on
the probe has global $E_8$ symmetry.

Now it is easy to consider smaller values of $N_f$.  In terms of string
theory Wilson lines break $E_8$ to a subgroup.  In the type I' language
this is represented by moving some of the background D8-branes {}from
the orientifold.  In terms of the theory on the 4-brane probe, this
corresponds to giving masses $m_i$ to some of the quarks.  Examining the
symmetry breaking pattern in string theory we see that for general
$N_f\le 7$ the enhanced symmetry is $E_{N_f+1}$.  For example, the
lowest non-trivial point has $SU(2)$ symmetry which occurs in string
theory at the self dual radius.  In the type I' theory there is no
background D8-brane at the orientifold, and in the 4-brane probe theory
this corresponds to a strong coupling fixed point with $N_f=0$.  $U(1)_I
\subset SU(2)$ is generated by the instanton charge.  For $g=\infty$ it
is enhanced to $SU(2)$.

\nref\witsmi{E. Witten, ``Small Instantons in String Theory,''
hep-th/9511030, \np{460}{1995}{541}.}
The Higgs branch of the theory on the brane is the moduli space of
space-time instantons.  For the ``trivial'' fixed points with $SU(N_f)$
or $SO(2N_f)$ symmetry, this fact is as in \refs{\witsmi,\bds,
\threedone}.  Our analysis shows that for the special points with
exceptional symmetries the Higgs branch again corresponds to the moduli
space of instantons, this time for the exceptional groups. 

\nref\ganoha{O. Ganor and A. Hanany, ``Small E(8) Instantons and
Tensionless Noncritical Strings,'' hep-th/9602120.}%
\nref\seiwit{N. Seiberg and E. Witten, ``Comments on String Dynamics in
Six Dimensions,'' hep-th/9603003.}%

These theories also appear in another way in string theory.  Consider a
compactification of the heterotic string on K3 with a small instanton.
Then, the six dimensional theory has enhanced gauge symmetry for
$SO(32)$ instantons \witsmi\ or is the mysterious tensionless string
theory for small $E_8$ instantons \refs{\ganoha,\seiwit}.  Upon further
compactification to five dimensions we find the five dimensional
theories discussed here.

\newsec{Strong coupling fixed points}

In the previous section we found new fixed points with global $E_n$
symmetry.  The parameters in these theories are background gauge fields
in the Cartan subalgebra of $E_n$, $m_i$ ($i=0,...,n-1$).  Turning on
$m_0$ we flow to the $SU(2)$ theory with $N_f=n-1$ flavors where $m_0$
is interpreted there as the (irrelevant parameter) $1\over g^2 $.  The
other parameters $m_i$ for $i=1,...,N_f$ are interpreted as the quark
masses in this IR theory.  Other deformations of the $E_n$ fixed point
theory are obtained by turning on some of $m_i$ for $i=p,..,N_f$.  The
theory then flows to the $E_p$ fixed point.  If we now turn on $m_0$ at
this fixed point, we flow to an $SU(2)$ gauge theory with $p-1$ flavors.

As we argued in the previous section, these $E_n$ fixed points have
$E_n$ global symmetry where $E_5=Spin(10)$, $E_4=SU(5)$,
$E_3=SU(3)\times SU(2)$, $E_2=SU(2)\times U(1)$ and $E_1=SU(2)$.  Their
the Higgs branches are isomorphic to the moduli spaces of $E_n$
instantons.  The Coulomb branches are ${\bf R}^+$.

One might ask whether all these theories are new.  In particular,
perhaps some of them are the same as known theories with $SU(2)$ or
$U(1)$ gauge symmetry.  These have global symmetries $SO(2N_f)$ or
$SU(N_f)$ and perhaps can be identified with some of the $E_n$ theories.
Furthermore, the Higgs branches are the same.  However, the Coulomb
branches of the $U(1)$ gauge theories are $\bf R$ while those of $E_n$
are ${\bf R}^+$.  This leaves only the $E_5$ theory as a potential
$SU(2)$ theory.  The relevant operator in $SU(2)$ with $N_f=5$ which
breaks the symmetry to $SU(5)\times U(1)$ takes it to a $U(1)$ theory
whose moduli space is $\bf R$.  In the $E_5$ theory such an operator
takes us to the $E_4$ theory whose Coulomb branch is ${\bf R}^+$.
Therefore, we conclude that all these theories are at interacting fixed
points of the renormalization group and are new field theories.

It is interesting to examine these theories in various dimensions.  In
six dimensions, there is a strange theory (small $E_8$ instantons) which
involves tensionless strings \refs{\ganoha,\seiwit}.  It is expected to
be a non-trivial field theory \seiwit\ with global $E_8$ symmetry.
Since the parameters are always background gauge superfields, and in six
dimensions there is no scalar in the vector multiplet, these six
dimensional theories do not have relevant operators which preserve the
super-Poincare symmetry.  

Upon compactification to five dimensions we find the $E_8$ theory
described here.  Unlike the situation in six dimensions, in five
dimensions we can describe the theory along the Coulomb branch by a
Lagrangian (in six dimensions there is a massless self-dual two form
which does not have a Lorentz invariant Lagrangian description).
Another difference is that in five dimensions we can perturb the theory
by relevant operators whose coefficients are $m_i$.  In terms of six
dimensional background gauge fields, these are Wilson lines around the
compact dimension which are scalars in five dimensions.  This allows us
to find the series of $E_n$ theories.

Upon further compactification to four dimensions we find new non-trivial
theories with $E_n$ global symmetry.  Now the parameters $m_i$ become
complex.  The Coulomb branch is one complex dimensional and the elliptic
curve describing the gauge coupling has an $E_n$ singularity.  The case
with $E_6$ in four dimensions was recently discussed in
\ref\minnem{J.A. Minahan and D. Nemeschanski, ``An $N=2$ Superconformal
Fixed Point with $E_6$ Global Symmetry,'' hep-th/9808047.}.

Upon further compactification to three dimensions the parameters $m_i$
become three vectors.  The Coulomb branch is a Hyper-Kahler manifold
with an $E_n$ singularity \refs{\threedone,\threedtwo, \intse}.  As
explained in \threedtwo, it is the four real dimensional auxiliary space
discussed in section 17 of
\ref\swtwo{N.~Seiberg and E.~Witten, \np{431}{1994}{484},
hep-th/9408099.}.

It is not known whether there are free field theories which flow to all
these non-trivial fixed points in more than three dimensions.  However,
in three dimensions such free field theories are known \intse.  The
existence of this UV free field theory which flows to the non-trivial
fixed points proves that they are indeed local quantum field theories
thus strengthening the claim (based on string theory) that all these
theories in all dimensions are local quantum field theories.

As we submitted this note, we received an interesting paper
\ref\ganor{O.J. Ganor, ``Toroidal Compactification of Heterotic 6D
Non-Critical Strings Down to Four Dimensions,'' hep-th/9608109.}
which addresses related issues from a different point of view.

\medskip
\centerline{\bf Acknowledgements}
This work was supported in part by DOE grant DE-FG02-96ER40559.  We
thank J. Polchinski, S. Shenker and E. Witten for helpful discussions.

\bigskip

\listrefs
\end